\documentclass[preprint,12pt]{elsarticle}

\usepackage{amssymb}
\usepackage{amsmath}
\usepackage{url}

\journal{Computers in Biology and Medicine}

\begin{document}

\begin{frontmatter}



\title{The segmentation ceiling: why explicit left-ventricular masks do not improve learned ejection-fraction regression}


\author[unlv]{Farshid Farhadi Khouzani\corref{cor1}}
\ead{farshid.farhadikhouzani@unlv.edu}
\author[unlv]{Paul La Plante}
\author[unlv]{Bryar Mustafa Shareef}
\author[unlv]{Laxmi Gewali}

\cortext[cor1]{Corresponding author. Department of Computer Science, University of
Nevada, Las Vegas, 4505 S. Maryland Pkwy, Las Vegas, NV 89154, USA.
Tel.: +1-702-357-1629; e-mail: farshid.farhadikhouzani@unlv.edu}

\affiliation[unlv]{organization={Department of Computer Science, University of Nevada, Las Vegas},
            city={Las Vegas},
            state={NV},
            country={USA}}

\begin{abstract}
Accurate estimation of left ventricular ejection fraction (EF) from
echocardiography is central to cardiovascular diagnosis and management, and deep
learning now enables automated EF prediction directly from echocardiographic
video. Because EF is clinically derived from left-ventricular (LV) volumes, a
widely held intuition is that providing a model with explicit LV segmentation
should improve prediction. We introduce a quantitative criterion, the
\emph{segmentation ceiling}, that makes this intuition testable: from the
definition of EF as a normalized difference of end-diastolic and end-systolic
volumes, we derive in closed form how per-frame segmentation area error propagates
into EF error, and thereby the segmentation accuracy a mask must reach before it
can improve on direct regression. Using the EchoNet-Dynamic dataset, a UniFormer-S
video backbone, and the empirically measured within-patient error correlation, the
criterion places the break-even at roughly 10\% per-frame area error, whereas a
representative segmenter operates at approximately 14\%, above the ceiling. The
criterion's prediction is confirmed by four independent strategies for injecting
segmentation or area information, namely (i) a predicted-mask input channel, (ii)
end-diastolic/end-systolic clip sampling, (iii) a per-bin area-consistency
objective, and (iv) an amplitude-consistency objective, none of which improves over
a raw-video baseline; ground-truth masks improve accuracy only through label
leakage. Having shown that input representation is not the limiting factor, we
identify generalization as the practical lever: an exponential moving average of
model weights combined with strong spatiotemporal augmentation attains a test
all-clips $R^2$ of 0.806 (MAE 4.08, RMSE 5.39 EF points) under a matched dense-clip
protocol, statistically comparable to a convolutional R(2+1)D baseline (0.811)
while substantially tightening the validation-to-test gap. Finally, we show that a
heteroscedastic $\beta$-NLL formulation yields informative, well-calibrated
per-prediction uncertainty, larger for clinically harder low-EF cases, where
post-hoc Monte-Carlo dropout does not. The segmentation ceiling provides a concrete
design criterion for when mask-guided EF estimation is worthwhile, together with a
simple, well-regularized, uncertainty-aware recipe for EF regression.
\end{abstract}




\end{frontmatter}



\section{Introduction}

Left ventricular ejection fraction (EF) is one of the most widely used
quantitative indicators of cardiac function. EF measures the fraction of blood
ejected from the left ventricle during systole and is routinely used in the
diagnosis, risk stratification, treatment selection, and longitudinal monitoring
of cardiovascular disease. In clinical practice, EF plays a central role in the
assessment of heart failure, cardiomyopathy, valvular disease, and treatment
response. Echocardiography is the most common imaging modality for EF assessment
because it is non-invasive, widely available, relatively inexpensive, and capable
of capturing cardiac motion in real time \cite{lang2015recommendations}.

Despite its clinical importance, EF estimation from echocardiography remains
challenging. Conventional EF measurement requires identifying end-diastolic and
end-systolic frames, tracing the left ventricular (LV) endocardial boundary, and
applying geometric assumptions such as Simpson's biplane method. These steps are
time-consuming and subject to inter- and intra-observer variability, particularly
when image quality is limited or ventricular boundaries are difficult to
delineate. Echocardiographic video additionally contains speckle noise, view
variability, patient-specific acquisition differences, and complex cardiac
motion, all of which complicate robust automated interpretation.

Deep learning has shown substantial promise for automated echocardiographic
analysis. Large-scale datasets such as EchoNet-Dynamic have enabled video-based
models that estimate EF directly from apical four-chamber echocardiography
\cite{ouyang2020video}, with convolutional spatiotemporal architectures such as
R(2+1)D \cite{tran2018closer} establishing strong baselines. Because EF is, by
clinical definition, computed from LV volumes via endocardial tracing, a natural
and widely held intuition is that providing a model with explicit LV
segmentation, whether as a preprocessing step, an auxiliary task, or an
additional input channel, should improve learned EF regression. A substantial
body of work accordingly couples LV segmentation with functional estimation.
However, whether explicit segmentation actually improves \emph{learned}
end-to-end EF regression, as opposed to merely matching the clinical computation,
has received surprisingly little direct scrutiny.

In this work we examine that assumption systematically and find that it does not
hold. We show, both analytically and empirically, that explicit LV segmentation
does not improve EF regression on EchoNet-Dynamic, and we explain
\emph{why} in quantitative terms. Our central observation is that EF is a
normalized \emph{difference} of end-diastolic and end-systolic volumes; this
subtraction amplifies per-frame segmentation error, so that a mask must be
extremely accurate before it carries more EF-relevant signal than the raw pixels
already provide. We formalize this as a ``segmentation ceiling'' and show that the required
per-frame area error (a break-even of roughly 10\% after accounting for the
measured ED/ES error correlation) lies beyond the reach of current
echocardiographic segmenters, which we measure to operate near 14\% area error. Consistent with this
analysis, multiple strategies for injecting segmentation or area information fail
to improve over a raw-video baseline, while ground-truth masks produce
artificially high accuracy attributable to label leakage, since the EF labels are
themselves derived from those traces.

Having established that the input representation is not the limiting factor, we
show that the practical bottleneck is generalization. A simple combination of an
exponential moving average (EMA) of model weights and strong spatiotemporal data
augmentation matches a convolutional R(2+1)D baseline under a carefully matched
dense-clip evaluation protocol, while substantially reducing the
validation-to-test generalization gap. We adopt UniFormer-S
\cite{li2022uniformer}, an efficient hybrid convolution-transformer video
backbone, as a representative modern architecture; our findings concern the role
of segmentation and regularization rather than any single architecture. Finally, we examine how per-prediction uncertainty should be obtained. We find that
post-hoc Monte-Carlo dropout \cite{gal2016dropout} yields uncertainty that does not
correlate with the model's actual error, and instead adopt a heteroscedastic
$\beta$-NLL formulation \cite{seitzer2022pitfalls} in which the network predicts
both the EF and its variance, producing informative, input-dependent uncertainty at
no additional inference cost.

The main contributions of this work are as follows:
\begin{itemize}
    \item We present a quantitative analysis, the \emph{segmentation ceiling},
    that explains why explicit LV segmentation cannot improve echocardiographic
    EF regression beyond a threshold set by per-frame area error, and we derive
    the accuracy a segmenter would need (per-frame area error $\lesssim$ 10\%,
    using the measured ED/ES error correlation) to provide any benefit.
    \item We empirically confirm the analysis through several
    segmentation- and area-guided strategies, including a predicted-mask input
    channel, end-diastolic/end-systolic clip sampling, and area-consistency
    auxiliary objectives, none of which improves over a raw-video baseline; we
    further show that ground-truth masks inflate accuracy through label leakage.
    \item We demonstrate that generalization, not input representation, is the
    practical bottleneck: EMA combined with strong augmentation attains accuracy
    statistically comparable to an R(2+1)D baseline under a matched dense-clip
    protocol while markedly reducing the validation-test gap.
    \item We show that the choice of uncertainty method matters: post-hoc
    Monte-Carlo dropout yields uncertainty uncorrelated with prediction error,
    whereas a heteroscedastic $\beta$-NLL formulation provides informative,
    input-dependent uncertainty that is larger for clinically harder low-EF cases.
    Throughout, we adopt a rigorous evaluation protocol (dense multi-clip inference
    with bootstrap confidence intervals).
\end{itemize}

\section{Related Work}

\subsection{Automated Ejection Fraction Estimation from Echocardiography}

Automated EF estimation from echocardiography has advanced rapidly, driven by the
clinical importance of EF and the availability of large datasets. Early deep
learning approaches used convolutional video architectures to learn spatiotemporal
representations directly from echocardiographic clips. The EchoNet-Dynamic
benchmark established a large-scale framework for EF prediction and LV segmentation
from apical four-chamber video \cite{ouyang2020video}, with convolutional
architectures such as R(2+1)D serving as strong baselines \cite{tran2018closer}.
Subsequent work narrowed the gap to expert agreement using attention-based and
hybrid designs; EchoCoTr, for example, combined convolution and self-attention and
reported an $R^2$ of approximately 0.82 on EchoNet-Dynamic
\cite{muhtaseb2022echocotr}, while earlier transformer pipelines applied
self-attention for ED/ES detection and EF computation \cite{reynaud2021ultrasound}.
Beyond point estimation, generative regression predicts a distribution over EF
values using conditional diffusion \cite{generative_regression_lvef_2026}, and
multi-task models jointly estimate EF alongside related measurements
\cite{zhou2026multiechonet}. Our study instead asks a more basic question: whether
explicit LV segmentation improves a strong video regressor at all.

\subsection{Transformer and Foundation Models for Echocardiography}

Transformer architectures are increasingly used in echocardiographic analysis for
their ability to model long-range spatiotemporal dependencies. We adopt UniFormer-S
\cite{li2022uniformer}, an efficient hybrid of convolution and self-attention, as a
representative modern backbone; our focus is the role of segmentation and
regularization rather than the architecture itself. Broader efforts include
foundation-model pretraining \cite{munim2026echojepa} and multi-modal
vision-language modeling for cardiovascular diagnosis \cite{pu2025mmvl}. Most
relevant to us, segmentation-free designs such as LV-STANet estimate LV functional
indices directly from video via spatial-temporal attention
\cite{batool2025lvstanet}, and automated strain-analysis pipelines support earlier
detection of dysfunction \cite{jiao2025digital}; these are consistent with our
finding that strong video models recover EF-relevant information directly from raw
pixels.

\subsection{Segmentation-Guided Echocardiography Analysis}

Because EF is clinically derived from LV volumes measured on segmented endocardial
contours, many pipelines incorporate segmentation, and considerable effort has gone
into improving echocardiographic segmentation itself, for instance through
motion-aware modules for sparse temporal labels \cite{hasan2026tam},
spatial-temporal consistency for semi-supervised segmentation \cite{guo2025stc}, and
graph-based reasoning over segmented structures for explainable EF
\cite{mokhtari2022echognn}. The relationship between segmentation quality and EF
accuracy has begun to receive direct attention. HSS-Net
\cite{efseg_hierarchical_2025} observes that segmentation models with high Dice
overlap can nonetheless yield poor EF estimates, and proposes a hierarchical
architecture to better preserve the frames and structures most relevant to EF.
Judge et al. \cite{uncprop_echo_2025} take a probabilistic view, sampling plausible
contours to propagate segmentation uncertainty into EF and fractional area change.
Both recognize that segmentation error is consequential for EF; they respond,
respectively, by improving the segmenter and by characterizing the resulting EF
uncertainty.

\subsection{Position of This Work}

Whether explicit segmentation actually benefits an end-to-end \emph{learned} EF
regressor, as distinct from supporting the classical volume-based computation, has
received little direct examination. Our work departs from prior segmentation-guided
approaches in three ways. First, we derive a closed-form, deterministic
\emph{break-even accuracy}: the per-frame area error below which a mask can improve
on direct regression at all. This differs from HSS-Net
\cite{efseg_hierarchical_2025}, which improves the segmenter in response to the
Dice--EF disconnect, and from Judge et al. \cite{uncprop_echo_2025}, who propagate
contour uncertainty into EF; to our knowledge, an explicit accuracy threshold and
its application to mask-guided learned regression have not appeared in prior
echocardiographic EF literature. Second, we confirm the criterion empirically:
several segmentation- and area-guided strategies fail to improve over a raw-video
baseline, while ground-truth masks inflate accuracy through label leakage. Third,
we show the practical bottleneck is generalization rather than input
representation, and that weight averaging with strong augmentation matches a
convolutional baseline while tightening the validation-to-test gap. We further find
that how uncertainty is obtained matters: post-hoc Monte-Carlo dropout yields
uncertainty uncorrelated with error, whereas a heteroscedastic $\beta$-NLL
formulation \cite{seitzer2022pitfalls} learns informative, input-dependent
uncertainty. We do not claim a new state-of-the-art EF method; rather, we clarify a
widespread assumption, quantify the ceiling it runs into, and offer a simple,
well-regularized, uncertainty-aware recipe.

\section{Dataset and Methods}

\subsection{Dataset}

\begin{figure}[t]
\centering
\includegraphics[width=0.8\linewidth]{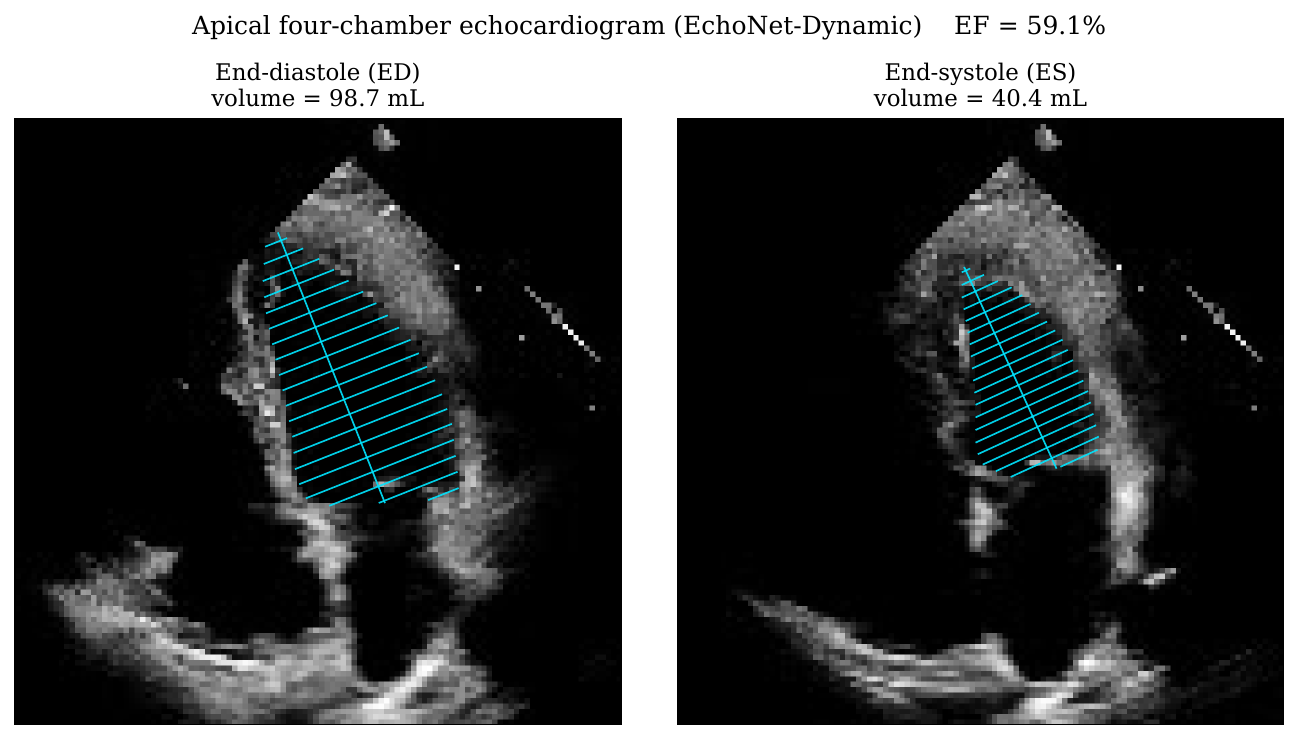}
\caption{Example apical four-chamber echocardiogram from EchoNet-Dynamic. Left:
end-diastolic frame (maximum left-ventricular volume). Right: end-systolic frame
(minimum volume). The expert left-ventricular tracing is overlaid.}
\label{fig:echo_example}
\end{figure}

We used the EchoNet-Dynamic dataset, a publicly available echocardiography video
dataset for automated cardiac function assessment \cite{ouyang2020video} (The dataset is publicly available for non-commercial research under a Stanford
Research Use Agreement\footnote{\url{https://echonet.github.io/dynamic/}}). The
dataset contains 10,030 apical four-chamber echocardiographic videos collected
during routine clinical care at Stanford University Hospital between 2016 and
2018. Each video is associated with a clinically measured left ventricular
ejection fraction (EF), provided as a continuous percentage value and used here as
the regression target. The released videos are standardized to $112 \times 112$
pixels and are cropped and masked to remove clinical text and patient
identifiers. EchoNet-Dynamic additionally provides expert left-ventricular
endocardial tracings at the end-diastolic (ED) and end-systolic (ES) frames of
each video, from which the reference EF is computed; we used these tracings to
construct the mask-guided input variants described below. Figure~\ref{fig:echo_example} shows a representative example, with the expert left-ventricular tracing at the end-diastolic and end-systolic frames. Following the standard
dataset split, our experiments used 7,465 videos for training, 1,288 for
validation, and 1,277 for testing. 

\subsection{Model Architecture}

We used UniFormer-S \cite{li2022uniformer} as the video backbone. UniFormer is a
hybrid architecture that applies local 3D convolution in shallow layers to reduce
spatiotemporal redundancy and global self-attention in deeper layers to capture
long-range dependencies, providing a favorable accuracy-efficiency trade-off for
video understanding. The backbone was initialized with Kinetics-400 pretrained
weights. We replaced the original classification head with a regression head
consisting of a dropout layer ($p = 0.5$) followed by a linear layer producing a
single scalar:
\begin{equation}
    \hat{y} = f_{\theta}(X),
\end{equation}
where $X$ is the input echocardiographic clip, $f_{\theta}$ the UniFormer-S
regressor, and $\hat{y}$ the predicted EF. The final-layer bias was initialized to
the approximate dataset-mean EF to stabilize early training. For the heteroscedastic variant described in Section~\ref{sec:uncertainty}, the
final linear layer produces two outputs rather than one, corresponding to the
predicted mean and log-variance.

We replaced all batch-normalization layers in the backbone with group normalization, which removes the dependence on batch statistics and yields stable optimization with the small batch sizes used here. To support the mask-guided input variants, which require a four-channel
input (three image channels plus one mask channel), we extended the first
patch-embedding convolution from three to four input channels by retaining the
pretrained weights for the image channels and initializing the additional channel
as the mean of the pretrained image-channel weights. Clips consisted of 36 frames
sampled at a temporal period of 4. As UniFormer-S expects a larger spatial
resolution than the native $112 \times 112$ frames, clips were resized to
$224 \times 224$ before being passed to the model.

\subsection{Input Preprocessing and Mask-Guided Variants}

For each video, frames were normalized using the training-set channel mean and
standard deviation. The mask-guided variants augment the three image channels with
a fourth channel encoding left-ventricular structure, allowing us to test whether
explicit segmentation improves EF regression. We considered three settings for
this channel, denoted by the mask source:
\begin{itemize}
    \item \textbf{Zero (baseline).} The fourth channel is set to zero, so the model
    receives only image information. This is our raw-video baseline and is
    architecturally identical to the mask-guided variants, isolating the effect of
    the mask content itself.
    \item \textbf{Ground truth.} The fourth channel contains the binary
    left-ventricular mask rasterized from the expert ED/ES tracings. Because the EF
    labels are derived from these same tracings, this setting is expected to leak
    target information and is included only as an upper reference, not as a
    deployable configuration.
    \item \textbf{Predicted.} The fourth channel contains the mask produced by a
    DeepLabV3 \cite{chen2017deeplabv3} segmentation network trained on
    EchoNet-Dynamic, which attains a Dice similarity coefficient of approximately
    0.92 and is representative of a realistic, deployable segmenter.
\end{itemize}

\subsection{The Segmentation Ceiling: A Closed-Form Criterion}
\label{sec:ceiling}

We now make the segmentation ceiling precise. Ejection fraction is defined as
\begin{equation}
    \mathrm{EF} = 1 - \frac{V_{\mathrm{ES}}}{V_{\mathrm{ED}}},
\end{equation}
where $V_{\mathrm{ED}}$ and $V_{\mathrm{ES}}$ are the end-diastolic and end-systolic
volumes. Suppose a segmenter estimates these volumes with multiplicative relative
errors $\varepsilon_{\mathrm{D}}$ and $\varepsilon_{\mathrm{S}}$, so that
$\hat{V}_{\mathrm{ED}} = V_{\mathrm{ED}}(1+\varepsilon_{\mathrm{D}})$ and
$\hat{V}_{\mathrm{ES}} = V_{\mathrm{ES}}(1+\varepsilon_{\mathrm{S}})$, where
$\varepsilon_{\mathrm{D}}, \varepsilon_{\mathrm{S}}$ are zero-mean with standard
deviation $\sigma_{\varepsilon}$ and within-patient correlation $\rho$. To first
order in the errors,
\begin{equation}
    \widehat{\mathrm{EF}}
    = 1 - \frac{V_{\mathrm{ES}}}{V_{\mathrm{ED}}}
      (1+\varepsilon_{\mathrm{S}})(1+\varepsilon_{\mathrm{D}})^{-1}
    \approx \mathrm{EF} - (1-\mathrm{EF})(\varepsilon_{\mathrm{S}} - \varepsilon_{\mathrm{D}}),
\end{equation}
so that the induced error in EF has standard deviation
\begin{equation}
    \sigma_{\mathrm{EF}}
    = (1-\mathrm{EF}) \, \sigma_{\varepsilon} \, \sqrt{2(1-\rho)}.
    \label{eq:ceiling}
\end{equation}

Equation~\eqref{eq:ceiling} is the segmentation ceiling in closed form, and it
makes three things explicit. First, the error is amplified by the factor
$(1-\mathrm{EF})$: because EF is a normalized difference of similar volumes, a given
relative volume error translates into an EF error scaled by how much blood remains
in the ventricle. Second, the amplification depends on the within-patient error
correlation through $\sqrt{2(1-\rho)}$: independent ED and ES errors ($\rho = 0$)
are maximally damaging, whereas a consistent multiplicative bias ($\rho = 1$)
cancels exactly, since a common scale factor divides out of the volume ratio.
Third, the criterion is quantitative. Equation~\eqref{eq:ceiling} yields a concrete break-even accuracy. For a
segmentation-derived EF to improve on a direct regressor, the induced error
$\sigma_{\mathrm{EF}}$ must fall below the regressor's own error. Under the
worst-case assumption of independent ED and ES errors ($\rho = 0$), setting
$\sigma_{\mathrm{EF}}$ equal to our model's mean absolute error of approximately
4.1 EF points gives a break-even of $\sigma_{\varepsilon} \approx 7\%$ per-frame
relative area error (at $\mathrm{EF} = 0.6$). Because real segmentation errors are
unlikely to be independent, we measured the within-patient correlation between the
ED and ES relative area errors of the DeepLabV3 segmenter directly, comparing the
predicted area against the expert-traced area at the end-diastolic and
end-systolic frames of each test video. This yields $\rho = 0.52$. With this
empirical correlation the break-even threshold relaxes to
$\sigma_{\varepsilon} \approx 10.5\%$. The same segmenter, however, operates at a
measured per-frame area error of $\sigma_{\varepsilon} = 13.8\%$, which induces an
EF error of approximately 5.4 EF points, above the regressor's own error
(Figure~\ref{fig:ceiling}). Thus, even after accounting for the realistic
correlation structure, a representative echocardiographic segmenter remains above
the segmentation ceiling, consistent with our empirical finding that predicted
masks do not improve, and in fact slightly degrade, EF regression.

\subsection{Segmentation- and Area-Guided Strategies}

To test the segmentation-ceiling prediction empirically, we evaluated four
strategies for injecting left-ventricular or area information into the regressor,
each compared against the architecturally identical zero-mask baseline.

\paragraph{Predicted-mask input channel.}
The four-channel input is populated with the DeepLabV3 predicted mask described
above, providing the model with explicit left-ventricular structure at every
frame.

\paragraph{End-diastolic/end-systolic clip sampling.}
Rather than sampling clips at arbitrary temporal positions, we biased clip
selection so that each training clip spans the annotated ED and ES frames,
ensuring that the frames most relevant to EF are present in every clip. This
tests whether explicitly exposing the model to the volumetric extremes improves
estimation.

\paragraph{Per-bin area-consistency auxiliary task.}
We added an auxiliary head that predicts the left-ventricular area at each
temporal position of the clip. Given backbone features of temporal length $T'$,
a spatial-pooling head produces a per-bin area sequence
$\mathbf{a} = (a_1, \dots, a_{T'})$, supervised at the bins corresponding to the
ED and ES frames using the areas derived from the expert tracings. The total
objective combines the EF regression loss with an area-supervision term and a
temporal-smoothness term,
\begin{equation}
    \mathcal{L} = \mathcal{L}_{\mathrm{EF}}
    + \lambda_{\mathrm{area}}\, \mathcal{L}_{\mathrm{area}}
    + \lambda_{\mathrm{smooth}}\, \mathcal{L}_{\mathrm{smooth}},
\end{equation}
encouraging the shared representation to encode the left-ventricular area
trajectory in addition to EF.

\paragraph{Amplitude-consistency auxiliary task.}
Because EF is determined by the amplitude of the area trajectory over a cardiac
cycle rather than by individual frame areas, we additionally derived a second EF
estimate directly from the predicted area sequence and required it to agree with
the EF label. To keep this estimate differentiable, we used soft maximum and
minimum operators over the area sequence,
\begin{equation}
    \widehat{\mathrm{EF}}_{\mathrm{area}}
    = \frac{\mathrm{softmax}_\beta(\mathbf{a}) - \mathrm{softmin}_\beta(\mathbf{a})}
           {\mathrm{softmax}_\beta(\mathbf{a})},
\end{equation}
where $\beta$ controls the sharpness of the soft extrema. The amplitude term
$\mathcal{L}_{\mathrm{amp}} = \lVert \widehat{\mathrm{EF}}_{\mathrm{area}} -
\mathrm{EF} \rVert^2$ was combined with the EF loss, a light smoothness term, and
a weak anchor that ties the predicted areas to the correct scale. Because the
area head is randomly initialized and its amplitude estimate is unreliable early
in training, the amplitude term was activated only after a short warmup period.

\subsection{Training, Weight Averaging, and Augmentation}

All models were trained for 45 epochs using discriminative learning rates, with a
lower rate for the pretrained backbone ($1\times10^{-5}$) and a higher rate for
the regression head ($1\times10^{-4}$), weight decay $1\times10^{-4}$, gradient
clipping, and a cosine learning-rate schedule with an eight-epoch linear warmup.
The optimization target was the mean squared error between predicted and reference
EF, except for the heteroscedastic variant, which was trained with the
$\beta$-NLL objective (Section~\ref{sec:uncertainty}).

To improve generalization we used two complementary techniques. First, we
maintained an exponential moving average (EMA) of the model weights with decay
$0.999$ \cite{tarvainen2017mean}; at each epoch we evaluated both the raw and EMA
weights on the validation set and retained whichever achieved the higher
validation $R^2$. Second, we applied stronger spatiotemporal data augmentation to
the training clips, comprising random horizontal flipping (which preserves EF
because area is reflection-invariant), random intensity scaling and brightness
shifts to emulate gain variation across ultrasound machines, and small random
rotations to emulate probe-angle variation. No augmentation was applied during
validation or testing.

\subsection{Evaluation Protocol}
\label{sec:eval}

We evaluated performance on both the validation and test sets and report results
under two inference settings. In the \emph{one-clip} setting, a single clip is
sampled per video. In the \emph{all-clips} setting, multiple clips per video are
evaluated and their predictions averaged to yield the video-level estimate. A
methodological subtlety arises here: the convolutional R(2+1)D baseline of
\cite{ouyang2020video} averages over \emph{every} temporally valid clip (dense
sampling), whereas a strided all-clips scheme averages over far fewer clips.
Because the number and spacing of averaged clips materially affects the reported
score, comparisons across models are valid only when this protocol is matched. We
therefore adopt the dense every-start all-clips protocol for all models reported
in our main comparison, ensuring that our results and the R(2+1)D baseline are
evaluated identically.

Regression performance was measured using the coefficient of determination
($R^2$), mean absolute error (MAE), and root mean squared error (RMSE):
\begin{equation}
    \mathrm{MAE} = \frac{1}{N}\sum_{i=1}^{N} |\hat{y}_i - y_i|,
\end{equation}
\begin{equation}
    \mathrm{RMSE} = \sqrt{\frac{1}{N}\sum_{i=1}^{N}(\hat{y}_i - y_i)^2}.
\end{equation}
We additionally evaluated threshold-based classification using the area under the
receiver operating characteristic curve (AUC) for clinically relevant EF
thresholds (EF $<35$, $<40$, $<45$, and $<50$). Confidence intervals for all
reported metrics were estimated using bootstrap resampling over the evaluation
set.

\subsection{Uncertainty Estimation}
\label{sec:uncertainty}

Per-prediction uncertainty is clinically valuable for automated functional
assessment. We consider two approaches: a widely used post-hoc method, and a
heteroscedastic model that predicts uncertainty directly.

\paragraph{Monte-Carlo dropout (baseline).}
Following \cite{gal2016dropout}, we keep the dropout layer of the regression head
($p = 0.5$) active at inference while the remainder of the network is held in
evaluation mode, and perform $T = 20$ stochastic forward passes per video. The
mean of the $T$ predictions is taken as the EF estimate and their standard
deviation as the predictive uncertainty. We note that the dropout layers within
the pretrained backbone are configured with a rate of zero and therefore introduce
no stochasticity; all variability originates from the head.

\paragraph{Heteroscedastic $\beta$-NLL regression.}
To obtain input-dependent uncertainty, we replace the scalar regression head with a
two-output head that predicts, for each video, a mean $\mu$ and a log-variance
$\log \sigma^2$, so that $\sigma^2 = \exp(\log \sigma^2)$ is positive by
construction. Assuming a Gaussian observation model, the negative log-likelihood
of a target $y$ is, up to a constant,
\begin{equation}
    \mathcal{L}_{\mathrm{NLL}}
    = \frac{1}{2}\left[ \frac{(y - \mu)^2}{\sigma^2} + \log \sigma^2 \right].
\end{equation}
Minimizing this loss couples the mean and the variance: the first term encourages
$\sigma^2$ to grow where the residual is large, while the second penalizes
unnecessarily large variance, so that at the optimum $\sigma$ reflects the
magnitude of the expected error for that input.

Optimizing $\mathcal{L}_{\mathrm{NLL}}$ directly is known to degrade the accuracy
of the mean, because the gradient of $\mu$ is scaled by $\sigma^{-2}$ and
therefore attenuated precisely on the difficult examples for which the model
predicts large variance. We therefore adopt the $\beta$-NLL objective
\cite{seitzer2022pitfalls}, which reweights each sample by a stop-gradient factor
$\left(\sigma^{2}\right)^{\beta}$:
\begin{equation}
    \mathcal{L}_{\beta\text{-NLL}}
    = \Big[ \left(\sigma^{2}\right)^{\beta} \Big]_{\perp}
      \cdot \mathcal{L}_{\mathrm{NLL}},
\end{equation}
where $[\,\cdot\,]_{\perp}$ denotes the stop-gradient operator. Setting $\beta = 0$
recovers plain NLL, while $\beta = 1$ fully compensates the $\sigma^{-2}$ scaling
so that the mean trains as under mean squared error. We use $\beta = 0.5$, the
value recommended in \cite{seitzer2022pitfalls} as the best trade-off between
predictive accuracy and uncertainty quality. For numerical stability the predicted
log-variance is clamped to a fixed range, and the variance-head weights are
initialized so that the initial $\sigma$ is on the scale of the expected EF error.

The mean $\mu$ is used for all point-estimate metrics ($R^2$, MAE, RMSE), so the
uncertainty head does not affect how accuracy is measured. We assess the quality of
the predicted $\sigma$ in two ways: by its correlation with the absolute prediction
error, which measures whether the uncertainty is informative, and by the empirical
coverage of the nominal $\mu \pm 2\sigma$ interval (approximately 95\%), which
measures calibration. Where a calibration correction is reported, a single variance-scaling factor is fit
on the validation set and applied unchanged to the test set; this rescales $\sigma$
without altering $\mu$ or the error--uncertainty ranking.

\section{Results}
\label{sec:results}

\subsection{Ejection Fraction Regression}

Table~\ref{tab:main} reports EF regression accuracy on the EchoNet-Dynamic test
set under the matched dense all-clips protocol (Section~\ref{sec:eval}). The
UniFormer-S configurations attain test $R^2$ values between 0.79 and 0.81, with
the EMA-plus-augmentation model reaching 0.806 (95\% CI 0.783--0.826), MAE 4.08,
and RMSE 5.39 EF points. All confidence intervals overlap with those of the
R(2+1)D baseline ($R^2 = 0.811$), indicating that the transformer and
convolutional models are statistically indistinguishable on this task once
evaluation is made directly comparable; the R(2+1)D baseline retains marginally
the best point estimates for MAE (4.01) and RMSE (5.32). We therefore do not claim
an improvement over the convolutional baseline. A modern video transformer
matches it, and reaches accuracy close to heavier transformer pipelines such as
EchoCoTr ($R^2 \approx 0.82$) \cite{muhtaseb2022echocotr}, with a simpler recipe.

\begin{figure}[t]
\centering
\includegraphics[width=0.85\linewidth]{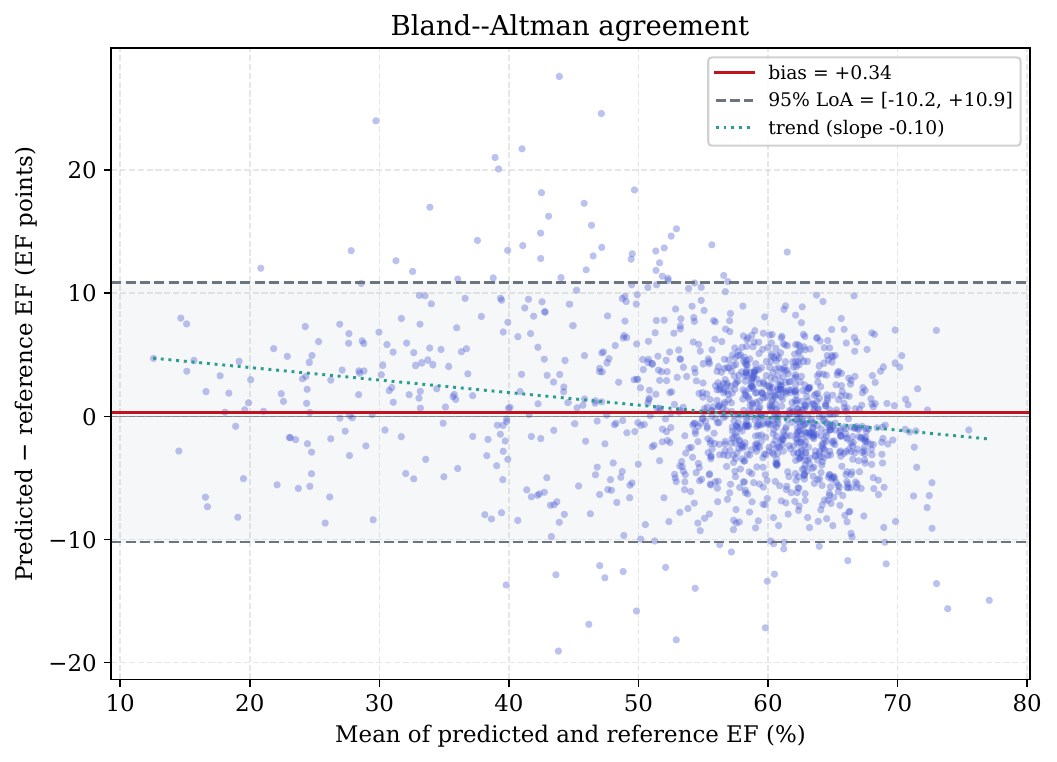}
\caption{Bland--Altman agreement between predicted and reference EF on the test
set, for the EMA-plus-augmentation model. The mean bias is $+0.34$ EF points, with
95\% limits of agreement of $[-10.2, +10.9]$ EF points; 94.7\% of predictions fall
within these limits. A slight negative trend (slope $-0.10$) indicates mild
regression toward the mean, with the model over-predicting low EF and
under-predicting high EF.}
\label{fig:bland_altman}
\end{figure}

Figure~\ref{fig:bland_altman} shows the corresponding Bland--Altman agreement
analysis: the model exhibits negligible systematic bias ($+0.34$ EF points) and
95\% limits of agreement of approximately $\pm 10$ EF points, with a slight
regression toward the mean at the extremes of the EF range.

\begin{table}[t]
\centering
\caption{EF regression on the EchoNet-Dynamic test set under the matched dense
all-clips protocol. Confidence intervals are 95\% bootstrap intervals. All test
$R^2$ intervals overlap, indicating statistically comparable performance.}
\label{tab:main}
\begin{tabular}{lcccc}
\hline
Model & Val $R^2$ & Test $R^2$ (95\% CI) & MAE & RMSE \\
\hline
R(2+1)D baseline \cite{ouyang2020video} & 0.817 & 0.811 (0.789--0.830) & 4.01 & 5.32 \\
UniFormer-S + EMA & 0.819 & 0.803 (0.781--0.822) & 4.11 & 5.43 \\
UniFormer-S + area-amplitude & 0.830 & 0.792 (0.768--0.813) & 4.24 & 5.58 \\
UniFormer-S + EMA + aug. & 0.815 & 0.806 (0.783--0.826) & 4.08 & 5.39 \\
\hline
\end{tabular}
\end{table}

\subsection{The Segmentation Ceiling and Mask-Guided Inputs}

\begin{figure}[t]
\centering
\includegraphics[width=0.85\linewidth]{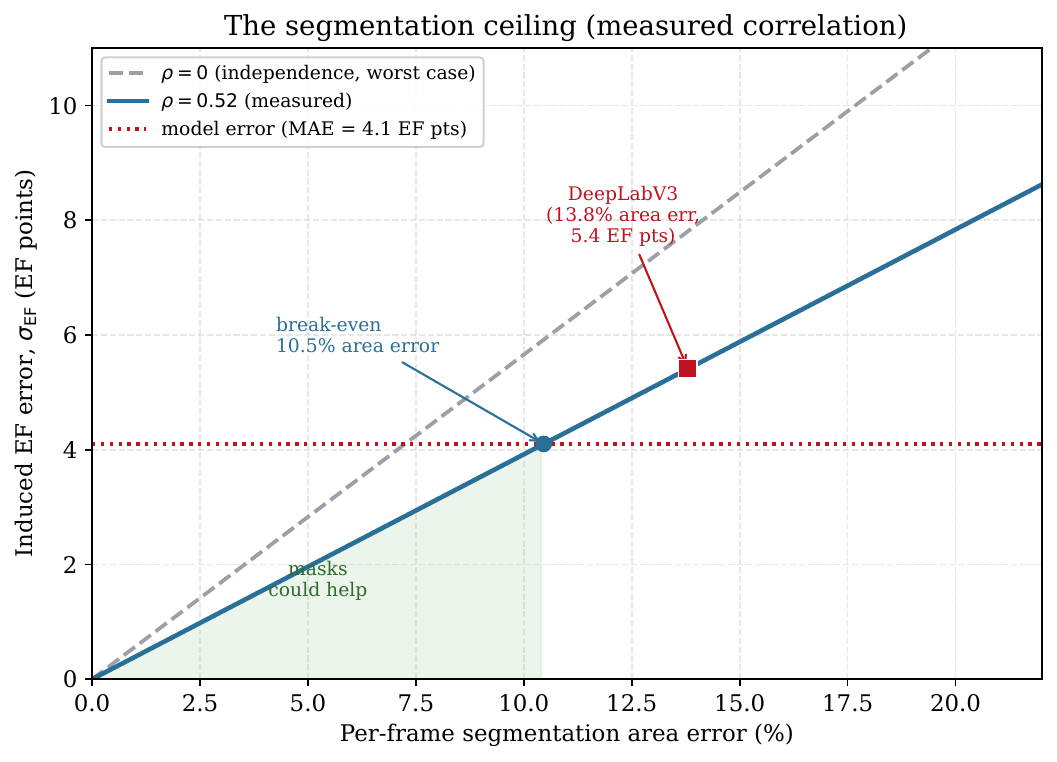}
\caption{The segmentation ceiling. Induced EF error $\sigma_{\mathrm{EF}}$ as a
function of per-frame segmentation area error (Eq.~\eqref{eq:ceiling}), for the
worst-case independence assumption ($\rho = 0$, dashed) and the measured
within-patient correlation ($\rho = 0.52$, solid). Masks improve on direct
regression only where the induced error falls below the model's own (red line, MAE
4.1 EF points), i.e.\ below a break-even area error of 10.5\%. A representative
DeepLabV3 segmenter (measured area error 13.8\%) lies above this ceiling.}
\label{fig:ceiling}
\end{figure}

Figure~\ref{fig:ceiling} summarizes the closed-form ceiling of
Section~\ref{sec:ceiling} with the measured error correlation. Using the empirical
$\rho = 0.52$, masks improve on direct regression only below a break-even area
error of 10.5\%, whereas the DeepLabV3 segmenter operates at 13.8\%, above the
ceiling. This predicts that realistic masks carry no net EF benefit, which the
input-channel experiments confirm.

This prediction is confirmed directly by the input-channel experiments. Supplying
the \emph{ground-truth} mask as a fourth channel yields a test $R^2$ of
approximately 0.97, but because the EF labels are derived from the same expert
tracings, this configuration leaks the target and is reported only as an upper
reference, not a deployable result. When the ground-truth mask is replaced by a
realistic DeepLabV3 predicted mask, performance does not merely fail to improve;
it \emph{drops} to 0.766 (95\% CI 0.737--0.790), below the 0.803 of the
architecturally identical zero-mask baseline (Table~\ref{tab:negatives}). The
imperfect mask thus injects a misleading area signal rather than useful structure.
We verified that the predicted masks are otherwise reasonable: predicted per-frame
areas correlate with ground-truth areas at $r \approx 0.96$, yet the residual
area error of roughly 14\% is sufficient to degrade EF estimation, exactly as the
ceiling analysis predicts. The limitation is therefore not poor localization but
the amplification of area error in the volume difference.


\subsection{Segmentation- and Area-Guided Strategies}

Table~\ref{tab:negatives} summarizes the strategies for injecting segmentation or
area information. None improves over the zero-mask baseline. The predicted-mask
channel and the end-diastolic/end-systolic (ED/ES) spanning clip-sampling variant
both fall clearly below baseline (0.766 and 0.773). The per-bin and amplitude
area-consistency objectives land just below baseline (0.791 and 0.792); although
they successfully induce the auxiliary head to encode the left-ventricular area
trajectory, with the amplitude-derived EF tracking the true EF during training,
they do not improve test accuracy. Notably, the amplitude objective \emph{widens}
the validation-to-test gap (0.038, versus 0.016 for EMA), indicating that the
auxiliary task adds capacity that fits the validation cohort without
transferring, rather than supplying EF information the regressor lacked.

\begin{table}[t]
\centering
\caption{Segmentation- and area-guided strategies versus the zero-mask EMA
baseline (dense all-clips test set). None improves over the baseline; the
predicted-mask and ED/ES-sampling variants fall clearly below it, and every
strategy enlarges the validation--test gap relative to the baseline.}
\label{tab:negatives}
\begin{tabular}{lcccc}
\hline
Configuration & Val $R^2$ & Test $R^2$ (95\% CI) & MAE & Val--test gap \\
\hline
Zero-mask baseline (EMA) & 0.819 & 0.803 (0.781--0.822) & 4.11 & 0.016 \\
\quad + predicted-mask channel & 0.790 & 0.766 (0.737--0.790) & 4.46 & 0.024 \\
\quad + ED/ES-spanning sampling & 0.799 & 0.773 (0.748--0.796) & 4.49 & 0.026 \\
\quad + per-bin area consistency & 0.818 & 0.791 (0.767--0.812) & 4.23 & 0.027 \\
\quad + amplitude consistency & 0.830 & 0.792 (0.768--0.813) & 4.24 & 0.038 \\
\hline
\end{tabular}
\end{table}

\subsection{Generalization: Weight Averaging and Augmentation}

Because input representation is not the limiting factor, we examined
generalization directly. Adding strong augmentation to the EMA model reduced
overfitting: the train-to-validation $R^2$ gap narrowed from 0.116 to 0.091, and
the validation-to-test gap narrowed from 0.016 to 0.009, the smallest among all
configurations and comparable to the 0.006 gap of the R(2+1)D baseline. Test
$R^2$ was essentially unchanged (0.803 to 0.806, overlapping intervals).
Augmentation thus brought validation and test performance into close agreement
without sacrificing accuracy, indicating that the residual gap reflects genuine
cohort difference rather than an evaluation artifact.

\subsection{Clinical Threshold Classification}

We assessed the model's ability to flag reduced EF at clinically relevant
thresholds. Table~\ref{tab:auc} reports AUC for the EF $<35$, $<40$, $<45$, and
$<50$ cutoffs for the EMA-plus-augmentation model. Discrimination is strong across
all thresholds, and validation and test AUCs track closely, in contrast to the
small val--test gap observed for $R^2$; this indicates that the threshold-level
classification generalizes cleanly even where continuous regression shows mild
cohort shift. The lowest threshold (EF $<35$) is based on relatively few positive
cases and is correspondingly less precise.

\begin{table}[t]
\centering
\caption{Threshold classification performance (AUC) at clinically relevant EF
cutoffs, for the UniFormer-S + EMA + augmentation model.}
\label{tab:auc}
\begin{tabular}{lcc}
\hline
EF threshold & Validation AUC & Test AUC \\
\hline
EF $< 35$ & 0.985 & 0.979 \\
EF $< 40$ & 0.981 & 0.978 \\
EF $< 45$ & 0.975 & 0.973 \\
EF $< 50$ & 0.959 & 0.954 \\
\hline
\end{tabular}
\end{table}

\subsection{Predictive Uncertainty}

Point predictions alone are of limited clinical value without an indication of their
reliability. We first evaluated Monte-Carlo dropout \cite{gal2016dropout} as a
baseline: its uncertainty estimates were essentially uninformative, with a
predictive standard deviation uncorrelated with the absolute error ($r \approx 0$)
and nominal intervals attaining far below their intended coverage, reflecting that
the network's dropout perturbations produce a nearly constant uncertainty across
inputs. This motivated the heteroscedastic $\beta$-NLL formulation, in which the
model predicts input-dependent uncertainty directly.

\begin{figure}[t]
\centering
\includegraphics[width=\linewidth]{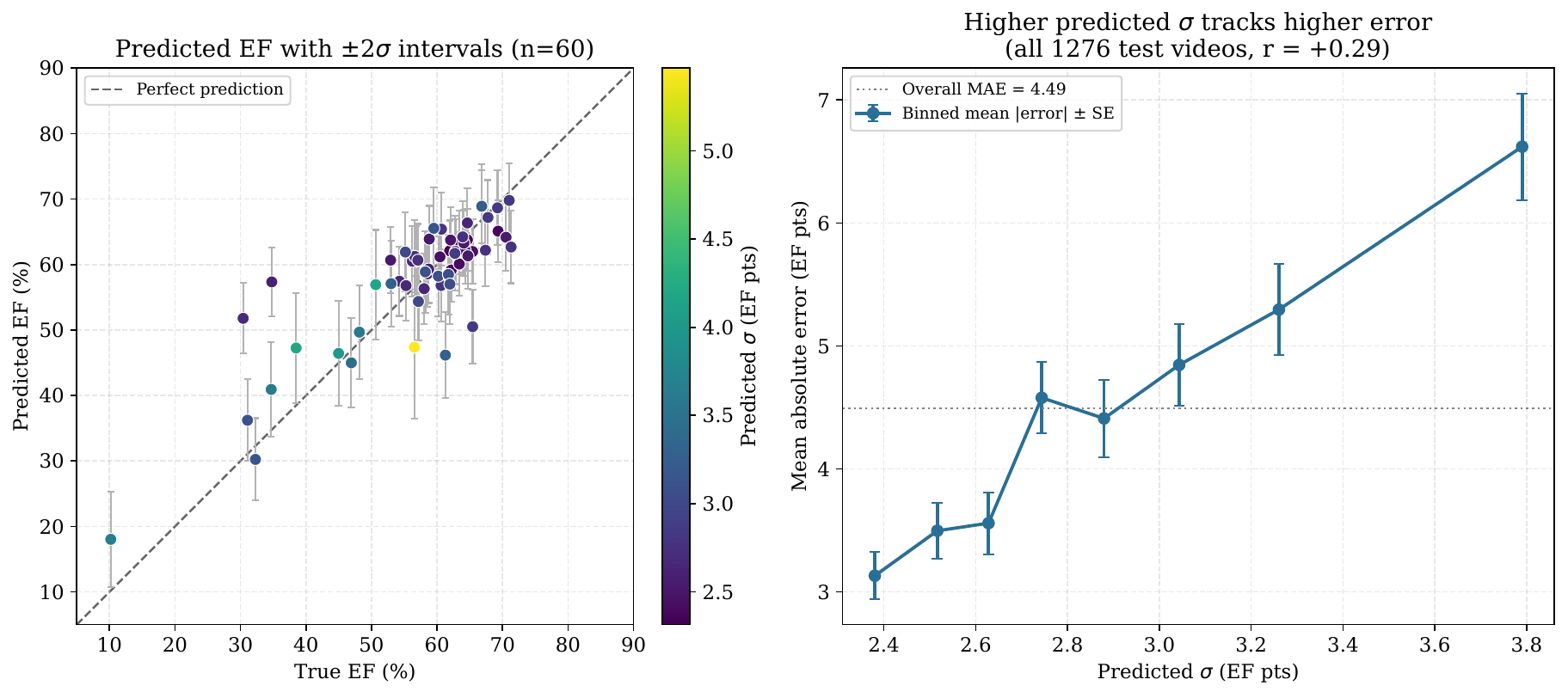}
\caption{Heteroscedastic $\beta$-NLL uncertainty on the test set. (a) Predicted
versus reference EF for a representative subset, with $\pm 2\sigma$ intervals
(approximately 95\%); point color encodes the predicted $\sigma$, showing that
uncertainty varies per case. (b) Mean absolute error as a function of predicted
$\sigma$, binned across all test videos: higher predicted uncertainty
corresponds to higher actual error ($r = 0.29$), confirming that the estimated
uncertainty is informative.}
\label{fig:uncertainty}
\end{figure}

The $\beta$-NLL model trades a modest amount of point accuracy for substantially
more informative uncertainty. Its dense all-clips test $R^2$ is 0.755 (95\% CI
0.723--0.782) with MAE 4.49, below the MSE-trained model (0.803, MAE 4.08); this
accuracy cost is comparable in magnitude to the differences reported among
configurations in Table~\ref{tab:negatives}, and is the price of adding the
uncertainty head. In return, the predicted uncertainty is markedly more useful,
as we now show. On the test set, the predicted standard deviation correlates with the
absolute error ($r = 0.29$), and the raw $\mu \pm 2\sigma$ intervals (nominal $\approx$95\%) attain 74\%
empirical coverage, indicating residual overconfidence but far better calibration
than the Monte-Carlo dropout baseline. Applying a single variance-scaling factor
fit on the validation set (Section~\ref{sec:uncertainty}) and applied unchanged to
the test set raises coverage to 93\%, close to the nominal level, without altering
the point estimates or the error--uncertainty ranking. Notably, the predicted uncertainty is larger for
lower-EF cases ($r = -0.39$ between $\sigma$ and EF), which are both
underrepresented in the dataset and clinically more difficult, indicating that the
model expresses greater uncertainty precisely where reduced cardiac function makes
assessment harder. Figure~\ref{fig:uncertainty} shows the predicted EF with
$\pm 2\sigma$ intervals, together with the relationship between predicted
uncertainty and observed error across the full test set.


\section{Discussion}

Our central finding is that explicit left-ventricular segmentation does not improve
learned EF regression, and that this is a predictable consequence of how EF is
defined rather than an incidental negative result. Because EF is a normalized
difference of end-diastolic and end-systolic volumes, per-frame area error is
amplified in the estimate, and the segmentation ceiling makes the required accuracy
quantitative and beyond the reach of current segmenters. Every strategy we tested is
consistent with this account: a realistic predicted-mask channel degraded accuracy
below the raw-video baseline, and ground-truth masks helped only through label
leakage.

The area-consistency experiments are especially informative because they separate
mechanism from benefit. The amplitude-consistency objective succeeded on its own
terms, learning a volume trajectory whose amplitude tracked true EF during training,
yet it produced no accuracy gain and slightly enlarged the validation-to-test gap.
Forcing the model to build an explicit internal volume representation thus added
nothing beyond what the regression head already extracts from raw pixels. This is
the clearest evidence that the limitation is not one of implementation or of the
``right'' representation; the available EF signal is bounded, and a strong video
model already approaches that bound.

A natural objection is that the area supervision could be made denser by segmenting
the intermediate frames. This does not circumvent the ceiling: the only available
source of intermediate-frame areas is a segmenter operating near 14\% area error, so
supervising the trajectory with these values injects the same signal already shown
unhelpful as an input channel, and because the amplitude depends on the difference
of extrema, that error is amplified rather than averaged out.

If input representation is not the bottleneck, generalization is. Weight averaging
with strong augmentation did not raise the accuracy ceiling but substantially
tightened the validation-to-test gap, and the residual gap reflects genuine cohort
differences between the EchoNet splits rather than an evaluation artifact. Future
gains are therefore more likely to come from larger, more diverse data and better
regularization than from architectural changes or segmentation-guided preprocessing.
For uncertainty, our results favor predicting it directly: Monte-Carlo dropout was
uncorrelated with error and overconfident, whereas the heteroscedastic $\beta$-NLL
model learns input-dependent uncertainty that correlates with error, is larger for
harder low-EF cases, and becomes well-calibrated after a single-parameter
recalibration. This uncertainty is informative rather than sharply discriminative,
and is best used to flag generally less-reliable predictions than to certify
individual cases. Clinically, the model discriminates reduced-EF cases well across
standard thresholds; combined with calibrated per-case uncertainty, this points
toward deployment in which automated EF estimates carry an explicit reliability
signal, even though continuous accuracy only matches the convolutional baseline.

\subsection{Limitations}

Several limitations qualify our conclusions. First, all experiments use a single
dataset from one institution (Stanford, 2016--2018), and the cohort shift we observe
even within EchoNet-Dynamic underscores that external validation is essential.
Second, we use the apical four-chamber view alone, whereas clinical EF is often
estimated from biplane measurements, so a single-view ceiling may differ from a
multi-view one. Third, the ceiling analysis is a first-order propagation summarized
at a representative EF and area-error scale; although we measured the ED/ES error
correlation directly, it still reduces a patient-dependent quantity to
representative values, and it is conducted on areas whereas EF is defined on
volumes, a simplification that is conservative since volume summation would amplify
area error further. Fourth, our conclusions concern a representative segmenter near
14\% area error; a substantially more accurate segmenter crossing the break-even
could in principle help, though none currently reaches that regime. Finally, the
uncertainty estimates derive from a single clip and a single model, their
correlation with error is modest, and the $\beta$-NLL point accuracy is slightly
below the MSE-trained model.

\subsection{Future Work}

Multi-view or biplane modeling may raise the accuracy ceiling by supplying
complementary geometric information, and larger, multi-site datasets would directly
address the cohort shift we identify as the dominant remaining source of error.
Richer uncertainty methods such as deep ensembles may yield more
error-discriminative estimates while retaining calibration. Finally, the
segmentation ceiling can serve as a design guideline, specifying the segmentation
accuracy mask-guided EF estimation would require to be worthwhile and thus when
investment in better segmentation would, and would not, pay off.

\section{Conclusion}

We revisited the widely held assumption that explicit left-ventricular segmentation
improves learned ejection-fraction regression, and showed, both analytically and
empirically, that it does not. Because EF is a normalized difference of
end-diastolic and end-systolic volumes, per-frame segmentation area error is
amplified in the estimate; our closed-form \emph{segmentation ceiling}, evaluated
with the measured within-patient error correlation, places the break-even at
roughly 10\% per-frame area error, whereas current segmenters operate near 14\%.
Consistent with this criterion, four strategies for injecting segmentation or area
information all fail to improve over a raw-video baseline, and ground-truth masks
help only through label leakage. With input representation ruled out, we identify
generalization as the practical lever: weight averaging with strong augmentation
matches a convolutional baseline while tightening the validation-to-test gap, and a
heteroscedastic $\beta$-NLL formulation supplies informative, well-calibrated
uncertainty where post-hoc dropout does not.

We do not claim a new state of the art. Our contribution is to explain
quantitatively why a natural and widely pursued direction saturates, to confirm
that explanation through controlled experiments, and to provide a concrete design
criterion: the segmentation accuracy that mask-guided EF estimation would require to
be worthwhile, and thus when investing in better segmentation would, and would not,
be expected to pay off.


\section*{Ethics statement}
This study uses the publicly available EchoNet-Dynamic dataset, which was released
in de-identified form under a Stanford University Research Use Agreement. No new
human-subjects data were collected, and no identifiable patient information was
accessed. Ethical approval for the original data collection is described in the
dataset's source publication.

\section*{Data availability}
This study uses the publicly available EchoNet-Dynamic dataset, accessible under a
Stanford Research Use Agreement at \url{https://echonet.github.io/dynamic/}. The
code used for the segmentation-ceiling analysis and the experiments in this paper
is available at \url{https://github.com/farhadcomp/echo-ef-ceiling}.

\section*{CRediT authorship contribution statement}
\textbf{Farshid Farhadi Khouzani:} Conceptualization, Methodology, Software,
Formal analysis, Investigation, Data curation, Writing -- original draft,
Visualization. \textbf{Paul La Plante:} Conceptualization, Methodology,
Supervision, Writing -- review \& editing. \textbf{Bryar Mustafa Shareef:}
Methodology, Writing -- review \& editing. \textbf{Laxmi Gewali:} Supervision,
Writing -- review \& editing.

\section*{Declaration of competing interest}
The authors declare that they have no known competing financial interests or
personal relationships that could have appeared to influence the work reported in
this paper.

\section*{Funding}
F.F.K. and P.L.P. are supported by Simons Foundation award number 00007127. This
work used the GPU cluster at the University of Nevada, Las Vegas. The funding
source had no role in the study design, in the collection, analysis, or
interpretation of data, in the writing of the manuscript, or in the decision to
submit it for publication.

\section*{Declaration of generative AI in the writing process}
During the preparation of this work the authors used a large language model to
assist with drafting, code development, and data analysis. After using this tool,
the authors reviewed and edited the content as needed and take full responsibility for the content of the publication.

\bibliographystyle{elsarticle-num}
\bibliography{references}
\end{document}